\documentstyle[psfig,emulateapj]{article}

\lefthead{Goldberg, Jones,  Hoyle, Rojas, Vogeley \& Blanton}
\righthead{The Void Galaxy Mass Function}

\begin{document}

\title{The Mass Function of Void Galaxies in the SDSS Data Release 2}

\author{David M. Goldberg$^1$, Timothy D. Jones$^1$, Fiona Hoyle$^1$, Randall
  R. Rojas$^{1,2}$, Michael S. Vogeley$^1$, \& Michael R. Blanton$^3$\\
$^1$ Department of Physics, Drexel University, Philadelphia, PA 19104;
  $^2$ Raytheon Corporation, Los Angeles, CA; $^3$ Department of
  Physics, New York University , New York, NY}

\begin{abstract}
We estimate the Mass Function of void galaxies in the second public
data release of the Sloan Digital Sky Survey from a sample of 1000
galaxies with local density contrasts of $\delta_v < -0.6$.  The
galaxy sample is split into ellipticals and spirals using a
color-S\'ersic index criterion.  We estimate the virial masses of
ellipticals using the measured spectral line-widths along with the
observed size.  Projection effects and uncertainties in halo
properties make mass estimates of spirals more difficult.  We use an
inversion of the Tully-Fisher relation to estimate the isothermal
rotational velocity, and introduce a scaling factor to estimate the
halo extent.  We then fit the measured mass function against a
theoretical Press-Schechter model, and find the distribution of
galaxies in voids appears to be nearly unbiased compared to the mass.
\end{abstract}

\section{Introduction}

Regions apparently devoid of galaxies (Kirshner et al. 1981) and
clusters (Einasto, Joeveer \& Saar 1980) were discovered in the early
1980's, and the existence of voids was confirmed by subsequent larger
surveys at a variety of wavelengths (de Lapparent, Geller, \& Huchra
1986; da Costa et al. 1988; Geller \& Huchra 1989; Davis et al. 1992;
Maurogordato et al. 1992; da Costa et al. 1994; see Rood 1988 and
references therein for a discussion of the history of void detection
and interpretation). Though their relative paucity has meant that void
galaxies have largely gone overlooked, they remain one of the best
probes of the effect of environment and cosmology on galaxy evolution,
and are perhaps one of the most intriguing new probes into our
understanding of structure formation.  Voids have been studied
statistically using techniques such as the void probability function
(Maurogordato \& Lachi\`eze-Rey 1987; Lachi\`eze-Rey, da Costa, \&
Maurogordato 1992; Vogeley et al. 1994; Croton et al. 2004; Hoyle \&
Vogeley 2004), found using void-finding techniques (Pellegrini, da
Costa \& de Carvalho 1989; Slezak, de Lapparent \& Bijaoui 1993;
El-Ad, Piran \& da Costa 1996; El-Ad, Piran and da Costa 1997;
M\"{u}ller et al. 2000; Plionis \& Basilakos 2002; Hoyle \& Vogeley
2002; 2004) and studied using semi-analytic or N-body simulations
(Mathis \& White 2002; Benson et al. 2003; Gottl\"ober et al. 2003).

Rojas et al. [2004a (photometric data), 2004b (spectroscopic data),
2004c (catalog)] have considered the properties of galaxies that
reside in extremely low density environments. They (Rojas et
al. 2004a) identify a sample of 10$^3$ void galaxies, i.e. galaxies
that are found in regions that have density contrast, $\delta_v\equiv
\delta \rho /\rho < -0.6$ detected using the Sloan Digital Sky Survey
(York et al. 2000, Stoughton et al. 2002, Abazajian et al. 2003,
Strauss et al. 2004).  The properties of galaxies in voids clearly
differ from those in higher-density regions, as seen in previous
studies of void galaxies that include examination of spectral and
photometric properties (Moody et al. 1987; Weistrop et al. 1995;
Popescu, Hopp \& Els\"{a}sser 1997, Grogin \& Geller 1999, 2000) and
HI content (Szomoru et al. 1996; Huchtmeier, Hopp, \& Kuhn
1997). Grogin \& Geller (1999, 2000) analyze a sample of 46 galaxies
in regions with density less than half of the mean density
(i.e. $\delta \rho/\rho < -0.5$) and find that these void galaxies are
bluer, of earlier type, and have a larger fraction of emission line
systems than galaxies in dense regions. Similarly, Rojas et
al. (2004a) find void galaxies are bluer, fainter and have
morphologies, as classified by their S\'ersic and concentration
indices, that more closely resemble late-type galaxies as compared to
galaxies that reside in higher density environments (wall galaxies).
Rojas et al. (2004b) also find that void galaxies have stronger
equivalent widths of H$\alpha$ and OII and have higher {\it specific}
star formation rates. Hoyle et al. (2003) measure the Luminosity
Function (hereafter LF) of these galaxies and find that the LF's of
the wall and void galaxies have different values of $M_r*$ (where
magnitudes are in SDSS bands unless stated otherwise) i.e. void
galaxies are fainter than wall galaxies, but the values of the faint
end slopes are very similar: $\alpha=-1.18\pm 0.13$ and
$\alpha=-1.19\pm0.07$ respectively. This suggests that voids are not
dominated by a large population of low luminosity galaxies.

An important question is whether voids are strongly anti-biased.  Do
they contain significant amounts of Dark Matter even though they are
largely devoid of light?  We would like to test this question by
determining the mass function of void galaxies and estimating their
local bias parameter.  The bias parameter, $b$ is defined as the ratio
of galaxy perturbations to the perturbations in the underlying dark
matter distribution.  For an unbiased distribution, $b=1$, the
density contrast of galaxies reflects the density contrast in dark matter.

Void regions also provide an important testbed for the overall picture
of galaxy formation, because Birkhoff's (1923) theorem suggests that
the behavior of structure growth within an under- or overdense region
will mimic that of a universe with the same mean properties.  Goldberg
\& Vogeley (2004) suggest a prescription to efficiently simulate the
growth of structure in voids by providing a mapping between the
cosmological parameters in the universe as a whole and the effective
parameters within the void region.  Thus, the formation and evolution
of galaxies within voids gives us the opportunity to test the
spherical collapse picture of halo formation (Press \& Schechter 1976)
within highly underdense regions.  Sheth \& van de Weygaert (2004)
explore a second level of excursion when an underdense void is nested
within a higher density region.

In this paper, we measure the void galaxy mass function in the Rojas
et al. (2004a) distant galaxy sample, and compare this to theoretical
models of void mass functions, in an attempt to understand the
environmental effects of low density regions on galaxy formation.  In
\S~\ref{sec:voidcat} we begin by introducing the SDSS void galaxy
catalog. Next, in \S~\ref{sec:massest} we discuss mass estimation of
the galaxies in this sample.  Because the SDSS does not include long
slit spectroscopy we do not have rotation curves for our sample.
Thus, we use an inversion of the Tully-Fisher relation to
statistically estimate the rotational velocities of our spiral sample.
In \S~\ref{sec:theory}, we present a theoretical basis for our
expectations of the mass function based on a Press-Schechter model
within an underdense region.  We then present the comparison of theory
with the measured mass function in \S~\ref{sec:results}, and find that
``typical'' void regions are consistent with an unbiased galaxy
formation picture.  We conclude with a discussion of future prospects.

\section{The Void Galaxy Catalog}
\label{sec:voidcat}

To obtain a sample of 10$^3$ void galaxies, we use data from the Sloan
Digital Sky Survey. The SDSS is a wide-field photometric and
spectroscopic survey. The completed survey will cover approximately
$10^{4}$ square degrees.  CCD imaging of 10$^8$ galaxies in five
colors and follow-up spectroscopy of 10$^6$ galaxies with $r<17.77$
will be obtained. York et al. (2000) provides an overview of the SDSS
and Stoughton et al. (2002) describes the early data release (EDR) and
details about the photometric and spectroscopic measurements.  Strauss
et al. (2004) describe the second data release (DR2).  Technical
articles providing details of the SDSS include descriptions of the
photometric camera (Gunn et al. 1998), photometric analysis (Lupton et
al. 2002), the photometric system (Fukugita et al. 1996; Smith et
al. 2002), the photometric monitor (Hogg et al. 2001), astrometric
calibration (Pier et al. 2003), selection of the galaxy spectroscopic
samples (Strauss et al. 2002; Eisenstein et al. 2001), and
spectroscopic tiling (Blanton et al. 2003a).  A thorough analysis of
possible systematic uncertainties in the galaxy samples is described
in Scranton et al. (2002). Galaxy photometry is k-corrected and
evolution corrected according to Blanton et al. (2003b).  We assume a
$\Omega_{\rm m}, \Omega_{\Lambda}$ = 0.3, 0.7 cosmology and Hubble's
constant $h = H_0/ 100$ km s$^{-1}$Mpc$^{-1}$ throughout.

Void galaxies are drawn from a sample referred to as {\tt sample10}
(Blanton et al. 2002), which is a subsample of the publicly available
DR2. This sample covers nearly 2000 deg$^2$ and contains 155,126
galaxies.  We use a nearest neighbor analysis to find galaxies that
reside in regions of density contrast $\delta \rho / \rho < -0.6$ as
measured on a scale of 7$h^{-1}$Mpc. These are the void galaxies. This
choice of density contrast and nomenclature is consistent with studies
of voids in more three-dimensional samples, in which individual void
structures are identified using an objective {\tt voidfinder}
algorithm (Hoyle \& Vogeley 2002, 2004). This definition finds voids
in the 2dFGRS, PSCz Survey and Updated Zwicky Catalog with typical
radii of 12.5$h^{-1}$Mpc. These voids fill 40\% of the Universe and
have mean density $\delta \rho / \rho < -0.9$. As expected, the
density around void galaxies ($\rho_{vg}$) is higher than the mean
density of a void ($\bar{\rho}_{void}$) because galaxies are clustered
and the few void galaxies tend to lie close to the edges of the voids.

Other techniques such as the method of El-Ad \& Piran (1997)
or use of tessellation techniques could also be used to find
void galaxies but currently the geometry of the SDSS does not allow
these techniques to be used as the SDSS is primarily comprised of thin
stripes which cannot wholly encompass the largest voids.  

The exact process of selecting the void galaxies is described in
detail in Rojas et al. (2004a). We provide a brief overview, as
follows: First, a volume limited sample with z$_{\rm max}=0.089$ is
constructed.  This is used to trace the distribution of the voids. Any
galaxy in the full flux-limited sample with redshift z$<$z$_{\rm max}$
that has less then three volume-limited neighbors in a sphere with
radius 7$h^{-1}$Mpc and which does not lie close to the edge of the
survey is considered a void galaxy. Galaxies with more than 3
neighbors are called wall galaxies.  Flux-limited galaxies that lie
close to the survey boundary are removed from either sample as it is
impossible to tell if a galaxy is a void galaxy or if its neighbors
have not yet been observed. This produces a sample of 1,010 void
galaxies and 12,732 wall galaxies. These void and wall galaxies have
redshifts in the range $0.034 < z < 0.089$ and have magnitudes in the
range $-22<$M$_{\rm r}<-17$ (Rojas et al. 2004a).

\section{Mass Estimation of Galaxies}
\label{sec:massest}

The mass function of galaxies is one of the most sensitive probes of
the effect of environment on the growth of structure.  The mass
function is directly related to the linear growth scale of structure
and the power spectrum of the CDM distribution.  One of the
complications in comparing a theoretical mass function to
observations, is that the simplest theories generally map the mass
function of Dark Matter halos, which are not directly observable.  In
the following section, we discuss methods for using observations to
estimate the halo masses.

\subsection{Classification of Morphologies}

Many properties of galaxies, such as color, luminosity and rotational
velocity vary with morphology.  The surface brightness profiles of the
different morphological types are found to vary predictably, with
spiral types begin more compact and ellipticals being more extended.
The surface brightness profiles of galaxies be well approximated by
the relation:
\begin{equation}
I(R) \propto \exp\left[-\left( \frac{R}{R_s}\right)^\frac{1}{n}\right] \ ,
\end{equation}
where $n$ is known as the S\'ersic (1968) index, such that $n=1$ for a
purely exponential disk, and $n=4$ for a de Vaucouleurs profile.

\begin{figure}[h]
\centerline{\psfig{figure=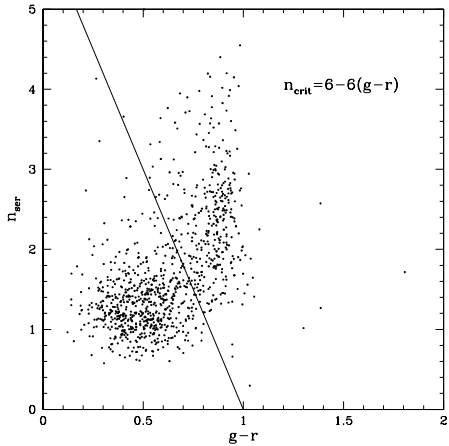,angle=0,height=2.5in}}
\caption{Void galaxies in S\'ersic index - color space.  A partition 
line of $n_{critical} = 6(1-(g-r))$ divides our sample.  
The masses of galaxies to the right of the partition line are
classified as ellipticals, the galaxies to the left of the partition line 
are classified as spirals.}
\label{fg:sersic}
\end{figure}

Strateva et al. (2001) found a correlation between morphology, color
($g-r$), and concentration. There is a bimodal distribution in
color-S\'ersic space.  Blanton et al. (2002) use $n < 1.5$ as the
selection criterion for Disks, and $n >3$ for Ellipticals.  Rojas et
al. (2004a) uses a $n_{critical} = 1.8$ cut to divide their sample
between Spiral and Elliptical types.  In this paper, we split the
sample in the color-n plane with $n_{critical} = 6(1-(g-r))$, as is
shown in Fig.~\ref{fg:sersic}. Galaxies
whose S\'ersic index fall below $n_{critical}$ are classified as
spirals, while galaxies whose S\'ersic index lie on or are above
$n_{critical}$ are classified as ellipticals.  

In our sample, we found 370 Ellipticals and 640 Spirals.  However,
since the sample is flux-limited and not volume-limited, in order to
estimate the fraction found in spirals we must weight by the volume in
which each galaxy could be detected.  Using this metric, we find that
82\% of void galaxies are Spirals, and 18\% are ellipticals.  This
extends the general morphology-density relation as found, for example,
by Postman \& Geller (1984).

\subsection{Mass Estimation of Ellipticals}

Elliptical galaxies have a much simpler structure than spirals, thus
simplifying the modeling of Ellipticals.  In this paper, we follow
Padmanabhan et al.(2004) and fit the circular velocity profile of
their model to N-body simulations.  They find a dynamical mass
estimate at $R_{50}$ of
\begin{equation}
M_{dyn,e} = \frac{(1.65\sigma)^2R_{50}}{G}\ ,
\label{eq:ell_dyn}
\end{equation}
where $R_{50}$ is the the circular half-light radius of a galaxy, and
$\sigma$ is the 1d velocity dispersion, as defined by the line-width
of H-$\alpha$.  We use this line both because of its strength, and
because of its completeness within our sample.  The SDSS obtains
spectra using a $3^{\prime\prime}$ fiber spectrograph, and thus it is
not possible to compute rotation curves or velocity profiles for the
galaxy sample.  However, Padmanabhan et al. constructed a composite
velocity profile for their sample of 29,469 SDSS galaxies and find a
nearly isothermal profile for early type galaxies.  Thus, the 1d
velocity width directly yields the temperature of the halo, greatly
simplifying the dynamical mass estimate of ellipticals.

Equation~(\ref{eq:ell_dyn}) only gives the mass out to the half-light
radius.  However, since Padmanabhan et al. (2004) assume that the halo
follows an NFW profile, the integrated mass is, in principle, well
defined.  Given only a velocity width and an apparent size, however,
we have too few parameters to uniquely determine an NFW profile (2
parameters), and a light profile (1 parameter) without implicitly
coupling the two.  Making an assumption that halo scales with the
apparent luminous scale, Padmanabhan et al. (2004) Fig. 10 suggests
that a factor of 4 will relate the total mass to that found in
equation~(\ref{eq:ell_dyn}).  We use this relation in our analysis
below.

\subsection{Mass Estimates of Spirals}

Ideally, we would like to directly measure the dynamical masses of
spirals as well as ellipticals.  However, the SDSS uses a fiber
spectrograph which does not allow us to directly probe the rotation
curves of spirals. However, we {\it can} estimate the rotation
velocity via an inversion of the Tully-Fisher (hereafter TF) relation.
Courteau (1997) finds a relatively small ($\sim 0.34$ mags) scatter in
the optical TF relation in the Lick $r$ band.  In terms of the most
relevant measures, Courteau's (1997) TF relation can be expressed as:
\begin{equation}
M^r_c=-6.17(\log v_{c,2.2}-2.5)-20.77
\end{equation}
where $v_{c,2.2}$ is the rotation radius at 2.2 times the scale length
(generally the peak velocity), and $M^r_c$ is the total integrated
magnitude in the Lick r filter.  The intrinsic scatter in
the TF relation can be inverted to produce a relatively small scatter
in the estimated rotational velocity, such that:
\begin{equation}
\frac{\sigma_{v_{c,2.2}^2}}{v_{c,2.2}^2}\simeq 0.26\ .
\end{equation}

The Lick r filter used by Courteau (1996) was based on the older Lick
Spinrad r filter.  However, Courteau (1996) demonstrates that their
photometry is effectively calibrated to the Gunn r system, to within
0.01 magnitudes for late-type galaxies.  Fukugita et al. (1995) show
that at low redshift the magnitude difference between the Gunn r
filter and the SDSS r' filter is $r'-r = -0.13$, which we adapt to
invert the Courteau (1997) Tully-Fisher relation.

Magnitudes are determined in the SDSS data set from Petrosian (1976)
radii, defined such that the mean circularly averaged flux within the
Petrosian radius is 5 times the flux over the annulus.  The SDSS
Petrosian magnitude (Blanton et al. 2001) is the measured magnitude
within 2 Petrosian radii.  For a simple exponential disk model (as is
likely to closely approximate the spiral sample), it can be shown that
the Petrosian magnitude should be $0.006$ higher than a theoretical
total magnitude.  This effect is significantly smaller than other
random and systematic errors which dominate this analysis.

Systematic uncertainty in the overall size of the galactic halos
dominates the uncertainty in mass estimates.  It is well known that
for galaxies the halo extends far beyond the visible limits of the
disk.  We thus define a Dark Halo Scale Factor, $S$, such that:
\begin{equation}
M_{dyn,s}=S\frac{v_{c,2.2}^2R_{90}}{G}\ .
\end{equation}
We have selected $R_{90}$ (the radius containing $90\%$ of the light)
to correspond to the visible extent of a disk, and thus, we expect
that the parameter $S$ will necessarily be larger than 1.  We can
approximate an upper bound on reasonable values for $S$ by considering
the Milky Way.  Given a mass of $M_{MW}\simeq 1.9\times 10^{12}
M_\odot$ (Wilkinson \& Evans 1999), an isothermal rotational velocity
of 220 km/s, and modeling the light from the Milky Way as a purely
exponential disk with a disk scale, $R_d=2.5kpc$ (Freudenreich 1998),
we can estimate a value of $R_{90}$ which would be observed for a
given projection angle.  Averaging over all angles, we find a limit of
about $S=35$.  Environment, however, is expected to strongly affect
this scale.  In \S~\ref{sec:results}, we show that it is possible to
use the Dark Halo Scale Factor as a free parameter in our fits, but
that high end values (such as that estimated from the Milky Way) give
a very poor fit between theory and observation.

We might reasonably wonder if $S$ is expected to be a constant
function of mass within a given environment.  Simple dimensional
analysis shows that given a canonical TF relation ($L\propto v^4$), an
assumption of constant central flux density, and a constant Mass-Light
ratio, $S$ will be constant.  In reality, we do not expect this to
perfectly model the mass function, and we thus leave the question of
the relation between dark matter extent and mass to future work.

\section{Press-Schechter Models of Galaxy Halos in Voids}
\label{sec:theory}

In the previous section, we have described our method for determining
the mass function of galaxies within cosmological voids.  We might
also be tempted to ask what mass function we might expect within a
void of a particular mean underdensity.  This question can be
addressed using the Press-Schechter (1974) formalism.  We use an
approach and notation following that of Mo \& White (1996; 2002), but within
the context of a constraint on the mean underdensity.

Our approach is quite similar to the work by Gottl\"ober et al. (2003;
following Sheth \& Tormen 2002), and for moderate values of the void
underdensity, the predicted mass functions are nearly identical.
However, there are several key differences.  First, Sheth \& Tormen
frame their discussion in terms of a barrier crossing under Brownian
motion.  Our formalism is wholly Bayesian.  Secondly, the constraint
placed by Gottl\"ober et al. is based on a polynomial fit to the
growth of an underdense perturbation.  We base our prior on a
numerical approach described in Goldberg \& Vogeley (2004).
Gottl\"ober et al. also note that, following Birkhoff's theorem, the
interior of the void may be treated as an isolated universe with mean
density given by the void density.  However, unlike Goldberg \&
Vogeley (2004), they do not correct the critical density of the void
(and thus the interior value of $\Omega$) to reflect the fact the
voids expand faster than the background universe, and thus have a
higher local value of the Hubble constant.  Finally, the formalism
allows an explicit dependence on the scale of the void.  Despite these
differences, even a in highly underdense region, $\delta_v\simeq
=-0.8$, the result presented below presents results which are
consistent with the Gottl\"ober et al. to within about 20\% up to
masses of $10^{11} M_\odot$.  Furthermore, as shown in Goldberg \&
Vogeley (2004) both estimates produce a satisfactory agreement in mass
function with both large-scale simulation, and the ``bottle universe''
simulations described in the same paper.

\subsection{Calculation of the Mass Function}

Consider a Gaussian random field of density perturbations,
$\delta(\vec{r})$, on which we apply an isotropic smoothing filter of
characteristic scale, $R$, $W(r;R)$, such that we have a smoothed
density field:
\begin{equation}
\delta(\vec{r}; R) \equiv \int W(|\vec{r}-\vec{r}|;R) \delta(\vec{r}')
d^3\vec{r}'
\end{equation}
As the new field is simply a linear sum of the underlying field, the
distribution of the smoothed field is also a Gaussian with mean zero
and a variance of
\begin{eqnarray}
\Delta^2(R)&\equiv&\langle \delta(\vec{r}; R)^2 \rangle \\ \nonumber
&=& \int P(k) |\hat{W}(k;R)|^2 d^3\vec{k}
\end{eqnarray}
where $\hat{W}(k;R)$ is the Fourier transform of the smoothing
kernel and $P(k)$ is the power spectrum of perturbations of the
underlying field:
\begin{equation}
P(k)=\langle \hat|\delta_{\vec{k}}^2|\rangle\ ,
\end{equation}
which is assumed to be isotropic.  

Now, let us further consider the covariance of a field which is
smoothed on two different scales.  
\begin{eqnarray}
V_{12} &\equiv&\langle \delta(\vec{r}; R_1) \delta(\vec{r}; R_2)
\rangle \\ \nonumber
& =&  \int P(k) \hat{W}(k;R_1) \hat{W}^\ast(k;R_2) d^3\vec{k}\ .
\label{eq:covardef}
\end{eqnarray}
Since the real-space window functions are spherically symmetric, the
Fourier space convolutions over the window functions on the void and
perturbation scale produce a real covariance.

Let us now consider a density perturbation, $\delta$, on a scale, $R$,
where all un-subscripted variables ($R$, $M$, $\delta$) are assumed to
be at the current epoch.  This perturbation arises in a larger region,
$R_v$, which has a mean {\it linear} underdensity, $\delta^L_v$.

We might imagine that at early times, a highly underdense void
satisfied the relationship, $\delta^L_{v,z}/D(z) < -1$.  In other
words, we would naively expect it to evolve to negative density today.
This is a natural limitation on linear theory and, thus, we define
$\delta_v$ as the underdensity that the void {\it would} have were it
allowed to linearly evolve indefinitely, a value which can be less
than -1.  Goldberg \& Vogeley (2004) derive an integral form for this
expression, and relate it to a simple parameter, $\eta$, such that:
\begin{equation}
\delta^L_{v}=\eta(\delta_v,\Omega_M,\Omega_\Lambda)\delta_0(\Omega_M,\Omega_\Lambda)
\end{equation}
where $\delta_v$ is the ``true'' underdensity of the void at the
present epoch, $\delta_v^L \le \delta_v$, $\Omega_M$ is the matter
density relative to critical, $\Omega_\Lambda$ is the cosmological
constant density relative to critical, and $\delta_0$ is the relative
structure growth factor as defined in Carroll, Press, \& Turner
(1992).

The joint probability density of finding a field with density
$\delta$ on scale $R$ and within a region of mean density $\delta_v^L$ on scale $R_v$
is thus a bivariate Gaussian with a covariance matrix:
\begin{equation}
V=\left(\begin{array}{cc}
\Delta^2 & V_{0v} \\
V_{0v} & \Delta_v^2
\end{array}
\right)
\end{equation}
where all terms in the covariance matrix are defined as parameters at
the present epoch.  However, since all perturbations are expected to
grow as $D(z)$ this relation can be readily modified (via a
substitution of $\delta_z=D(z)\delta$ and so on) at all times.

It can thus be shown that for a bivariate Gaussian with the covariance
matrix as above, 
\begin{eqnarray}
f(\delta ; R | \delta^L_v ; R_v) &=& {\cal N}
\left(\delta_v
\frac{V_{0v}}{\Delta_v^2},
\sqrt{\Delta^2-\frac{V_{0v}^2}{\Delta_v^2}}\right)\\
&\equiv&{\cal N}[\mu(R),\sigma(R)]
\label{eq:bivariate}
\end{eqnarray}
where ${\cal N}(\mu,\sigma)$ represents the normalized Gaussian distribution
function.  For brevity, we will henceforth not explicitly state the prior
condition $\delta_v^L, R_v$.

Defining:
\begin{equation}
x\equiv \frac{\delta-\mu}{\sigma}
\end{equation}
equation~\ref{eq:bivariate} simply becomes $f(x)={\cal N}(0,1)$.

A principal result of the spherical collapse model (Press \& Schechter
1974; Peebles 1980) is that (in an $\Omega_M=1$ universe) any region
for which the smoothed density is greater than $\delta_c > 1.69$ has
reached a maximum expansion in the linear approximation and has begun
to collapse.  However, this critical overdensity is similar in most
cosmologies.  

Bond et al. (1991) show that the cumulative probability of finding a
collapsing/collapsed mass exceeding $M$ at redshift $z$ is thus:
\begin{equation}
F(M,z)={\rm erfc}\left(\frac{\nu}{\sqrt{2}}\right)
\end{equation}
where the factor of 2 in front stems from normalization of the mass
distribution (for a pedagogical discussion see Padmanabhan 2000).
Note that R and M are equivalent measures of a perturbation, such
that:
\begin{equation}
M=\frac{4\pi}{3}\rho_0 R^3\ ,
\end{equation}
where $\rho_0$ is the mean density of the universe at present.  Finally,
\begin{equation}
\nu\equiv \frac{\delta_c/D(z)-\mu}{\sigma}
\end{equation}

Relating the perturbation distribution to a mass function, we get:
\begin{eqnarray}
n(M,z)dM&=&-\frac{dF}{dM} \frac{\rho_0}{M}(1+\delta_{v,z})dM\\ \nonumber
&=&\sqrt{\frac{2}{\pi}}
\exp\left(-\frac{\nu^2}{2}\right)(1+\delta_{v,z})\frac{\rho_0}{M}\frac{d\nu}{dM}dM
\end{eqnarray}
where the $1+\delta_{v,z}$ term normalizes the comoving volume (set by
the background universe) to the physical volume at any given time, and
$\nu$ is implicitly a function of redshift.

This model was derived using spherical collapse approximations.
However, following Sheth, Mo \& Tormen (2001), we can incorporate an
elliptical collapse model:
\begin{equation}
n(M,z)dM= A\left( 1+\frac{1}{\nu'^{2q}}\right) \sqrt{\frac{2}{\pi}}
\exp\left(-\frac{\nu'^2}{2}\right)(1+\delta_{v,z})\frac{\rho_0}{M}\frac{d\nu'}{dM}dM
\end{equation}
where $\nu'=\sqrt{a}\nu$, $a=0.707$, $A=0.322$, and $q=0.3$.
Simulations (e.g. Jenkins et al. 2001) suggest that the ellipsoid
model produces a good fit to galaxy abundances.  We use the
ellipsoidal distribution function throughout the forgoing analysis.
 
\subsection{The Growth of Galaxies in Voids}

Before moving on to the observed mass function, and a comparison with
theory, we discuss briefly the implications of the growth of void
galaxies in the model above.  It is clear that with less matter
available to grow galaxies, void regions must necessarily contain far
fewer of them.  Indeed, this is how voids are identified.  Moreover,
if we were to identify perturbations of equal amplitude at early times
in a void and in the background, the background perturbation would
grow into a visible galaxy faster, and would have a much larger final
mass than the void galaxy.

\begin{figure}[h]
\centerline{\psfig{figure=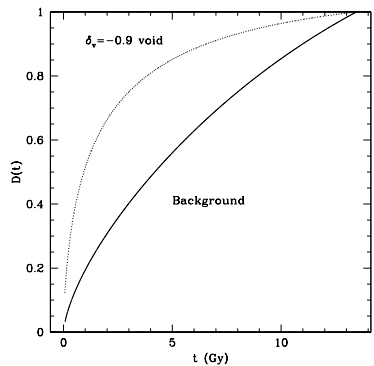,angle=0,height=2in}}
\caption{The normalized linear growth parameter within a background,
  $\Omega_M=0.3$ flat cosmology, and within a $\delta_v=-0.9$ void.
  Note that the void growth parameter achieves a higher fraction of
  the present value in the void than in the background for all points
  in the past.}
\label{fg:times}
\end{figure}

However, we wish to ask the converse: given an observed void galaxy or
galaxy distribution {\it today}, what is a likely formation history?
Goldberg \& Vogeley (2004) discuss the evolution of the linear growth
parameter, $D(z)$ in some detail.  As Fig.~\ref{fg:times}
illustrates, the growth parameter achieves a higher fraction of its
present value at early times in voids than in the background universe.
This argument is similar to the ones accompanying the normalization of
$\sigma_8$ in open versus flat cosmologies, stemming from the
well-known result that perturbations ``freeze-out'' at $z\simeq
\Omega_M^{-1}-1$ in low-density universes.

In other words, we would expect that given two galaxies of equal
masses, one identified in the field, and one in a void, the void
galaxy would have been more fully formed (in terms of its present
mass) at earlier times.

\begin{figure}
\centerline{\psfig{figure=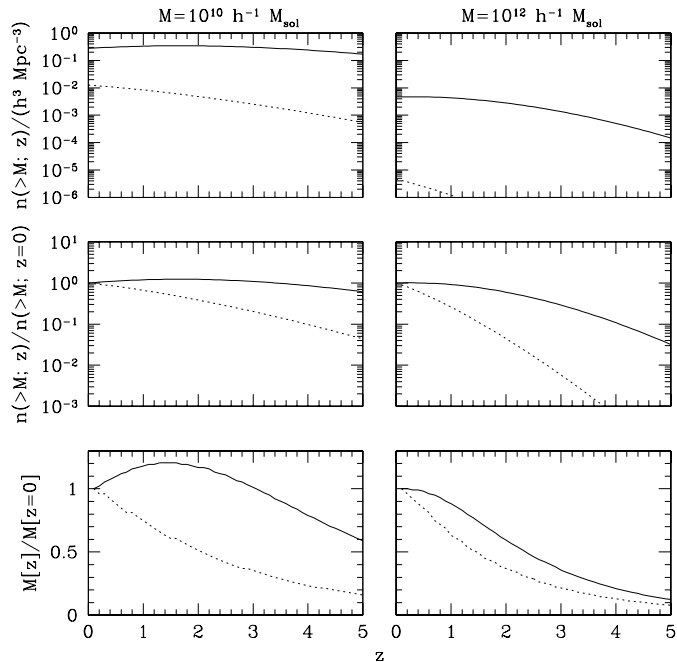,angle=0,height=4in}}
\caption{We plot the evolution of a theoretical galaxy mass function
  in a $\delta_v=-0.9$ void (dotted lines), and in the background
  (solid).  For each, we compute the evolution of the $N(>10^{10}
  h^{-1}M_\odot)$ (left) and $N(>10^{12}h^{-1}M_\odot)$ (right)
  distribution.  The top set of panels shows the evolution of the
  cumulative mass function for each fiducial mass.  These are
  normalized to comoving volume.  The middle panels show the mass
  function normalized to the present.  The bottom panels show the
  evolution of a halo mass (described in the text) of $10^{10}$ and
  $10^{12} h^{-1} M_\odot$ galaxies at earlier epochs.  }
\label{fg:evolve}
\end{figure}

Another way of looking at this is in terms of the evolution of the
mass function.  In Fig.~\ref{fg:evolve} we show the time-dependence of
the $M > 10^{10}$ and $M > 10^{12} h^{-1}M_\odot$ slice through the
mass function in both voids (dotted), and in the background cosmology
(solid).  We do not use the $1+\delta_v$ volume correction, so that at
each epoch we have the same comoving volume.  At first glance,
Fig.~\ref{fg:evolve} gives an interpretation of structure growth that
is the opposite of what we conclude from Fig.~\ref{fg:times}.  For
both intermediate and high mass galaxies, we find a larger fractional
increase in the number density at recent epochs in void regions than
in the background.  This effect is due to the fact that the mass
function is much steeper at high masses for voids than for the
background.  As a result, a small growth in structure results in the
production of (fractionally) many more galaxies at high masses.

Finally, we can ask, ``How much matter has any {\it given} galaxy
accreted since a previous epoch?''  We may set a density threshold,
and ask, for a given epoch, for what mass is the cumulative mass
function equal to that density?  The bottom panel in
Fig.~\ref{fg:evolve} shows the evolution of this function for
$10^{10}$ and $10^{12} h^{-1}M_\odot$.  The fact that in the
background a $10^{10}h^{-1}M_\odot$ galaxy achieves a maximum mass
around $z\simeq 1.5$ suggests that a significant fraction of this mass
accretion is in the form of mergers.  This simple picture suggests
that for the few high mass galaxies in voids, we would expect a
relatively high fraction of recent mass accretion.  This is consistent
with the result found by Rojas et al. (2004b) who found a
relatively high specific star formation rate in voids compared to wall
regions.

\section{Results: The Void Mass Function in the SDSS}
\label{sec:results}

\begin{figure}[h]
\centerline{\psfig{figure=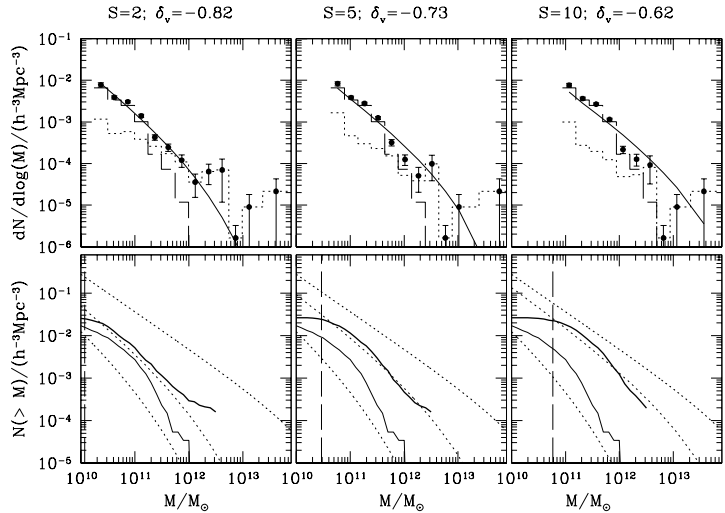,angle=0,height=4in}}
\caption{A comparison between theoretical and estimated dynamical mass
functions of the void galaxies in the SDSS Data Release 2.  In each,
we vary the Dark Halo Scale Factor, $S$, such that $S=2,5,10$ from
left to right.  {\bf Top panels:} The differential mass functions,
with errorbars within bins given by Poisson errors.  Random errors of
$26\%$ in $v^2$ are applied by convolving (flattening) the model
distribution function.  The solid line represents the best fit DM halo
model of $\delta_v=\{-0.82,-0.73,-0.62\}$.  No errors are given for
the (unknown) systematic uncertainties arising from the halo extent.
In each, the long-dashed histogram shows the distribution of Spirals,
and the short-dashed histogram shows the distribution of Ellipticals.
{\bf Bottom panels:} The cumulative mass functions.  The horizontal
line represents the mass detection limit for each assumed scale
factor.  The bold line represents the measured dynamical mass
function.  The thin line represents the mass function obtained from an
assumed mass-stellar mass ratio of 3 from the stellar mass estimates
of Kauffmann et al.  Note that the inferred masses of ellipticals are
the same in all three panels, and only the spiral mass estimates vary
with S. Dotted lines show analytic DM mass functions, for
$\delta_v=-0.9$ (bottom), $\delta_v=\{-0.82,-0.73,-0.62\}$ (middle),
and $\delta_v=0$ (top). }
\label{fg:massfunc}
\end{figure}

In Figure~\ref{fg:massfunc} we show the estimated SDSS void galaxy
mass function for Dark Matter Scale Factors, $S=$2, 5 and 10.  For
comparison, we also show the expected Press-Schechter mass functions
for $\delta_v=-0.9$ (observed mean galaxy underdensity) and
$\delta_v=0$ (background) environments, and determine the best value
of $\delta_v$ for each value of $S$, by matching the cumulative number
density of objects at 4 times the mass detection limit.  

We note that the minimum value of $R_{90}$ observed in the spiral
sample is approximately 2 kpc, and that the maximum observable
absolute magnitude is -18. Combining these results, we find a minimum
detectable mass of $S\times 6\times 10^{9} h^{-1} M_\odot$.  This mass
detection limit is shown by a vertical dashed line in the bottom
panels of Fig.~\ref{fg:massfunc}.  The mass detection limit is
determined by relating the flux-limited minimum velocity estimate with
a fit relation between size and luminosity.  It should be noted that
the mass function does not flatten out considerably beyond this limit.
In other words, we do not see a deficit of low mass/low luminosity
galaxies in voids beyond what is expected by Press-Schechter analysis
and our detection limit.  In addition to plotting the mass function
for all galaxies, we subdivide our sample into ellipticals and
spirals in Fig.~\ref{fg:massfunc}.  Ellipticals clearly consitute the
high-mass end of the spectrum, and  their distribution tends to be
much flatter than the distribution of spirals, which dominate at low
and intermediate mass.

For comparison, we plot a mass function estimated from the stellar
mass distribution given by Kauffmann et al. (2003) from the SDSS First
Data Release (Abazajian et al. 2003).  To turn these into total mass
distributions, we assume a constant mass-stellar mass ratio of 3, as
estimated for ellipticals by Padmanabhan et al. 2004.  The
mass-stellar mass ratio of spirals is expected to be larger than for
ellipticals, even if the baryon ratio for both is the same, since
spirals are expected to be more gas rich.  It is clear, however, that
this simple estimate of the mass function does not produce a good fit
to the slope of the Press-Schechter mass functions.

For our ``dynamical'' mass estimates, lower values of S appear to
produce a better fit to the shape of a Press-Schechter mass function.
This is confirmed via a series of $\chi^2$ tests.  For $S=2,5,10$, the
fits to $\delta_v=-0.82$, $\delta_v=-0.73$, and $\delta_v=-0.62$,
respectively, produce $\chi^2$ per degree of freedom of $3.3$, $8.2$
and $14.7$.  Note that we only model this fit out to masses of
$5\times 10^{12}M_\odot$.  Beyond that, all three models produce at
most 1 galaxy per bin, in each case, an Elliptical.  Since these
galaxy masses are somewhat unreasonably large, it may be that we have
simply underestimated the uncertainty in velocity measurement for the
Ellipticals.

Since in all cases we measure $\chi^2 > 1$, it is clear that despite
an excellent ``chi-by-eye,'' we have not correctly characterized
either our uncertainties or our mass measurements.  One of the most
likely culprits is that the halo extent parameter, $S$, is an explicit
function of mass.  Given the other uncertainties in our measurements,
it is unrealistically optimistic to try to claim a functional form of
this term with any confidence.

\begin{figure}
\centerline{\psfig{figure=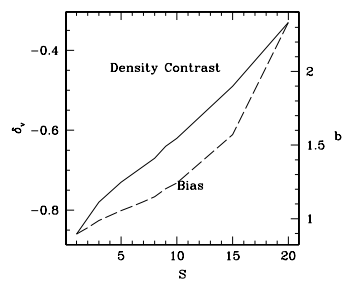,angle=0,height=2in}}
\caption{The relationship between best fit density contrast and halo
  extent parameter.  Note that for relatively compact halos (relative
  to the luminous distribution) we find a fairly unbiased distribution
  for low values of $S$ and an only modestly (positively) biased
  distribution for larger values of $S$. }
\label{fg:bias}
\end{figure}

Since a given value of $S$ implies an underdensity in Dark Matter for
the ensemble of voids, we show, in Fig.~\ref{fg:bias} the explicit
relationship between these two terms.  Moreover, a simple calculation
of the ``typical'' galaxy density within our selected void sample
(number of observed galaxies, divided by total void volume), yields
$\delta_{v,gal}=-0.77$.  Since lower values of $S$ are generally
preferred (from a $\chi^2$ point of view), we that voids may, in fact,
be nearly unbiased tracers of dark matter.   Even for larger values of
$S=10$, we find $b=1.24$, consistent with typical bias relations found
in the universe at large.

\section{Discussion}
\label{sec:discuss}

There remain several untested assumptions in this study.  For example,
though compact halos are most consistent with our theoretical mass
function models, Press-Schechter has not been tested against
observations in very low density void regions.  Moreover, we assumed
that the TF relation in the field would necessarily hold for void
galaxies as well.  However, as Rojas et al. (2004a) has shown, the
photometric properties of void galaxies differ from those of wall
galaxies, and thus we would not be surprised to find a different TF
relation.

Future prospects for work in this direction include more systematic
estimates of the rotation curves of void spirals.  Ideally, followup
observations using long-slit spectroscopy could potentially yield a TF
relation for voids which differs significantly from wall regions.

Additionally, as the numbers of known void galaxies increase,
galaxy-galaxy gravitational lensing will become a potentially powerful
probe for measuring mass profiles and halo extents.  Since
the separation between a lens and a source galaxy is much larger than
the scale of a void ($\Delta z\simeq 0.5$ in many cases) an isolated
void galaxy lens may still have many potential sources to lens.  

\acknowledgements

DMG acknowledges support from NSF grant AST-0205080.  MSV acknowledges
support from NSF grant AST-0071201.  We thank Michael Strauss for
useful suggestions.


\begin{thebibliography}{DUM}
\bibitem{} Abazajian, K., et al., 2003, AJ, 128, 502
\bibitem{} Benson, A. J., Hoyle, F., Torres, F \& Vogeley, M. S.,  2003, MNRAS, 340, 160
\bibitem{} Birkhoff, G.D. 1923, ``Relativity and Modern Physics,''
  (Cambridge, Mass: Harvard University Press)
\bibitem{} Blanton, M. R., et al., 2001, AJ, 121, 2358
\bibitem{} Blanton, M. R., et al., 2002, ApJ, in press
\bibitem{} Blanton, M. R., et al., 2003a, AJ, 125, 2276
\bibitem{} Blanton, M. R., et al., 2003b, AJ, 125, 2348
\bibitem{} Bond, J.R., Cole, S., Efstathiou, G., \& Kaiser, N., 1991, ApJ 379, 440
\bibitem{} Carroll, S. M.,  Press, W. H. \&  Turner, E. L., 1992, ARA\&A, 30, 499
\bibitem{} Courteau, S., 1996, ApJS, 103, 363
\bibitem{} Courteau, S., 1997, AJ, 114, 2402
\bibitem{} Croton, D. J., et al. 2004, MNRAS submitted, astro-ph/0401406
\bibitem{} da Costa, L.N., et al., 1988, ApJ, 327, 544
\bibitem{} da Costa, L.N., et al., 1994, ApJ, 424, 1
\bibitem{} Davis, M., Huchra, J.P., Latham, D.W., \& Tonry, J., 1992, ApJ, 253, 423
\bibitem{} de Lapparent, V., Geller, M.J., \& Huchra, J.P., 1986, ApJ, 302, L1
\bibitem{} Einasto, J., Joeveer, M., \& Saar, E.,  1980, MNRAS 193, 353
\bibitem{} Eisenstein, D. J. et al. 2001, AJ, 122, 2267
\bibitem{} El-Ad, H., \& Piran, T. 1996, ApJ, 462, 13
\bibitem{} El-Ad, H., \& Piran, T. 1997, ApJ, 491, 421
\bibitem{} El-Ad, H., Piran, T., \& da Costa, L.N., 1997, MNRAS, 287,
  790
\bibitem{} Freudenreich, H.T., 1998, ApJ 492, 495
\bibitem{} Fukugita, M.,  Shimasaku, K. \& Ichikawa, T., 1995, PASP, 107, 945
\bibitem{} Fukugita, M., Ichikawa, T., Gunn, J.E., Doi, M., Shimasaku,
K. \& Schneider, D. P., 1996, AJ, 111, 1748
\bibitem{} Geller, M. J., \& Huchra, J. P., 1989, Science, 246, 857
\bibitem{} Goldberg, D.M. \& Vogeley, M.S., 2004, ApJ, 605, 1
\bibitem{} Gottl\"{o}ber, S., {\L}okas, E. L., Klypin, A., and
  Hoffman, Y., 2003, MNRAS 344, 715
\bibitem{} Grogin, N., \& Geller, M.J. 1999, AJ, 118, 2561
\bibitem{} Grogin, N., \& Geller, M.J. 2000, AJ, 119, 32
\bibitem{} Gunn, J.E., et al. 1998, ApJ, 116, 3040
\bibitem{} Hogg, D.W., Finkbeiner, D.P., Schlegel, D.J., \& Gunn,
J.E. 2001, AJ, 122, 2129
\bibitem{} Hoyle, F. \& Vogeley, M. S., 2002, ApJ, 566, 641
\bibitem{} Hoyle, F., Rojas, R., Vogeley, M.S., \& Brinkmann, J., 2003
ApJ submitted, astro-ph/0309728
\bibitem{} Hoyle, F., \& Vogeley, M.S., 2004, ApJ Accepted,
  astro-ph/0312533
\bibitem[]{}
Huchtmeier, W. K., Hopp, U., \& Kuhn, B. 1997, A\&A, 319, 67 
\bibitem{} Jenkins, A et al.,  2001, MNRAS, 321, 372
\bibitem{} Kauffmann, G. et al. 2003, MNRAS, 341, 54
\bibitem{} Kirshner, R. P., Oemler, A. Jr., Schechter, P. L.,
Shectman, S. A. 1981, ApJ, 248, 57
\bibitem{} Lachi\`eze-Rey, M., da Costa, L. N. \& Maurogordato, S.,
1992, ApJ, 399, 10
\bibitem{} Lupton, R. H., Ivezic, Z., Gunn, J. E., Knapp, G., Strauss, M. \&
Yasuda, N., 2002, SPIE, 4836, 350
\bibitem{} Mathis, H. \& White, S.D.M., 2002, MNRAS 337, 1193
\bibitem{} Maurogordato, S. \& Lachi\`eze-Rey, M., 1987, ApJ, 320, 13
\bibitem{} Maurogordato, S., Schaeffer, R. \& da Costa, L.N., 1992, ApJ, 390, 17
\bibitem{} Mo, H.J. \& White, S.D.M., 1996, MNRAS 282, 347
\bibitem{} Mo, H.J. \& White, S.D.M., 2002, MNRAS, in press
\bibitem{} Moody, J. W., Kirshner, R. P., MacAlpine, G. M. \&
Gregory, S. A., 1987, ApJ, 314, 33
\bibitem{} M\"{u}ller, V., Arabi-Bidgoli, S., Einasto, J., \& Tucker,
D. 2000, MNRAS, 318, 280
\bibitem{} Padmanabhan, N. {\it et al.}, 2004, NewA 9, 329
\bibitem{} Peebles, P.J.E., 1980, ``The Large-Scale Structure of the
  Universe,'' (Princeton: Princeton University Press).
\bibitem{} Pellegrini, P.S., da Costa, L.N., \& de Carvalho, R.R., 1989, 339, 595
\bibitem{} Petrosian, V., 1976, ApJ, 209,  L1
\bibitem{} Pier, J., et al. 2003, AJ, 125, 1559
\bibitem{} Plionis, M., \& Basilakos, S. 2002, MNRAS, 330, 399
\bibitem[]{} Popescu, C., Hopp, U., \& Els\"{a}sser, H., 1997, A\&A, 325, 881 
\bibitem{} Postman, M. \& Geller, M.J., 1984, ApJ 281, 95
\bibitem{} Press, W.H. \& Schechter, P., 1974, ApJ, 187, 425
\bibitem{} Rojas, R., Vogeley, M.S., Hoyle, F. \& Brinkmann, J.,
  2004a, accepted in ApJ, astro-ph/0307274
\bibitem{} Rojas, R., et al., 2004b, submitted to ApJ, astro-ph/0409074
\bibitem{} Rojas, R., et al., 2004c, submitted to ApJ.
\bibitem{} Rood, H.J., 1988, ARA\&A, 26, 245
\bibitem{} Scranton, R. et al. 2002, ApJ, 579, 48
\bibitem{} S\'ersic, J. L., 1968, Atlas de Galaxias Australes, Observatorio
Astron\'omico, Cordoba
\bibitem{} Sheth, R.K., Mo, H.J., \& Tormen, G., 2001, MNRAS 323, 1
\bibitem{} Sheth, R.K. \& Tormen, G. ,2002, MNRAS, 329, 61
\bibitem{} Sheth, R.K. \& van de Weygaert, R., 2004, MNRAS 350, 517
\bibitem{} Slezak, E., de Lapparent, V., \& Bijaoui, A., 1993, ApJ, 409, 517
\bibitem{} Smith, J.A., et al. 2002, AJ, 123, 2121
\bibitem{} Stoughton, C. et al. 2002, AJ, 123, 485
\bibitem{} Strateva, I., et al. 2001, AJ, 122, 1861
\bibitem{} Strauss, M., et al. 2002, AJ, 124, 1810
\bibitem{} Strauss, M., et al. 2004, AJ, 128, 2004
\bibitem[]{} Szomoru, A., van Gorkom, J. H., Gregg, M. D. \& Strauss, M. A., 1996, AJ, 111, 2150 
\bibitem{} Vogeley, M.S., Geller, M.J., Park, C., \& Huchra, J.P., 1994, ApJ, 108, 745
\bibitem{} Wilkinson, M.I. \& Evans, N.W., 1999,  MNRAS 310, 645
\bibitem[]{} Weistrop, D., Hintzen, P., Liu, C., Lowenthal, J., Cheng, K. P.,
Oliversen, R., Brown, L. \& Woodgate, B., 1995, AJ, 109, 981 
\bibitem{} York, D. G., et al. 2000, AJ, 120, 1579
\end{thebibliography}
\end{document}